\documentclass[11pt]{cernrep}
\usepackage{graphicx}
\usepackage{here}
\def\slashchar#1{\setbox0=\hbox{$#1$}
   \dimen0=\wd0 \setbox1=\hbox{/} \dimen1=\wd1
   \ifdim\dimen0>\dimen1 \rlap{\hbox to \dimen0{\hfil/\hfil}} #1
   \else  \rlap{\hbox to \dimen1{\hfil$#1$\hfil}} / \fi}

\begin{document}
\title{THE SPECTRAL QUARK MODEL AND LIGHT-CONE
                                PHENOMENOLOGY\thanks{Presented at
                                ``Light Cone Workshop: HADRONS AND
                                BEYOND'', 5h-9th August 2003,
                                University of Durham (U.K.)}}
                                \author{\underline{Enrique Ruiz
                                Arriola}$^1$ and Wojciech
                                Broniowski$^2$} \institute{
                                $^1$Departamento de F\'{\i}sica
                                Moderna, Universidad de Granada,
                                E-18071 Granada, Spain \\ $^2$ The
                                H. Niewodniczanski Institute of
                                Nuclear Physics, Polish Academy of
                                Sciences, PL-31342 Cracow, Poland}
                                \maketitle
\begin{abstract}
Chiral quark models offer a practical and simple tool to describe
covariantly both low and high energy phenomenology in combination with
QCD evolution. This can be done in full harmony with chiral symmetry
and electromagnetic gauge invariance. We review the recently proposed
spectral quark model where all these constraints are implemented.
\end{abstract}

\section{INTRODUCTION}

Light cone hadron properties provide a non-trivial playground to test
ideas and methods on genuine non-perturbative QCD phenomenology. As we
have heard from S. Brodsky in this workshop this information is
conveniently encoded in LC wave functions~\cite{Brodsky:2001wx}. They
are measured through high energy inclusive, such as Deep Inelastic
Scattering (DIS), or exclusive processes, which with the help of QCD
renormalization group evolution can be related to low-energy matrix
elements.  In the particularly interesting case of the pion one
expects chiral symmetry to impose certain constraints. The specific and
general way how this is done seems unclear at the moment, because by
asking about the quark content of a hadron we are actually in a
situation where both hadrons and quarks coexist as degrees of freedom,
so standard effective field theory tools like Chiral Perturbation
Theory~\cite{Gasser:1983yg} do not apply directly beyond predicting
finite pion mass corrections.  On the other hand, fundamental lattice
approaches naturally formulated in Euclidean space can only obtain low
moments of structure functions (see e.g. the talk of P. E. L. Rakow
and Ref.~\cite{Gockeler:2002ek}) and distribution amplitudes, which
are not testable directly through experiment. As B. van de Sande
presented in his talk, QCD transverse lattice approaches can be directly
formulated on the light-cone and although promising results for the
pion and other mesons exist already~\cite{Dalley:2002nj} (for a review
see e.g. Ref.~\cite{Burkardt:2001jg}) they do not easily accommodate
for chiral symmetry. The Schwinger-Dyson approach in its Euclidean
formulation and its implications~\cite{Maris:2003vk} has been reviewed
by C. D. Roberts.

In order to gather some theoretical insight on the role of chiral
symmetry in high energy inclusive~\cite{Davidson:2001cc} or
exclusive~\cite{RuizArriola:2002bp} processes we remain at the more
modest level of the so-called quiral quark models reviewed in
Ref.~\cite{RuizArriola:2002wr}. This generic name stands for
phenomenological relativistic quantum field theories where chiral
symmetry is spontaneously broken incorporating Goldstone's theorem.
One of the advantages of such an approach is that it is covariant, so
we avoid from the beginning the difficult marriage between light-cone
quantization and chiral symmetry. In this talk we review salient
features of recent work done by us on a new chiral quark model: the
spectral quark model
(SQM)~\cite{RuizArriola:2001rr,RuizArriola:2003bs,Broniowski:2003rp}.
This model shares many good aspects of other chiral models,
particularly NJL type models, but improves others at a very low
computational cost and, up to now, brings very encouraging
phenomenological success.


QCD is a theory of quarks and gluons as elementary degrees of freedom,
so both contribute to the total momentum carried by a hadron. The
separation between gluons and valence and sea quarks, however, is both
scale and renormalization scheme dependent which can be worked out
explicitly within perturbative QCD. On the other hand, chiral quark
models which are supposedly a low energy approximation to QCD,
obviously saturate the momentum sum rule, since they contain no
explicit gluonic degrees of freedom. This poses the natural question:
is there any scale in QCD at which the momentum fraction carried by
the (valence) quarks is exactly one ?

According to the phenomenological QCD analysis undertaken by the
Durham group over a decade ago~\cite{Sutton:1991ay}, the momentum fraction
carried by the valence quarks is 0.47 $\%$ at the scale $\mu^2 = 4
{\rm GeV}^2 $,  e.g., for $\pi^+$,
\begin{eqnarray} 
V_1(\mu) = \langle x \left( u_\pi - \bar u_\pi + \bar d_\pi - d_\pi \right)
\rangle = 0.47 \pm 0.02 \qquad {\rm at} \qquad \mu^2 = 4 {\rm GeV}^2
\; .
\end{eqnarray} 
where $u_\pi $, $\bar u_\pi $ , $d_\pi $ and $\bar d_\pi $ are parton
distribution functions (PDF) in the pion.  Assuming LO perturbative
QCD evolution one gets
\begin{eqnarray}
\frac{ V_1 (\mu) } { V_1 (\mu_0) } = \left( \frac{\alpha(\mu)}
{\alpha(\mu_0) } \right)^{\gamma_1^{\rm NS} / 2 \beta_0 } \quad ,
\qquad  \alpha(\mu)= \left( \frac{4 \pi}{\beta_0 } \right) \frac1{\log (\mu^2
 / \Lambda_{\rm QCD}^2 )}
\end{eqnarray} 
where $ \gamma_1^{\rm NS} / 2 \beta_0 = 32/81 $ for $N_F=N_c=3$. We
take for concreteness $\Lambda_{\rm QCD}=226~{\rm MeV} $, which for
$\mu = 2 {\rm GeV} $ yields $\alpha=0.32 $. Downward LO evolution
yields that for the reference scale, $\mu_0$,
\begin{eqnarray}
V_1 (\mu_0)  =1 , \qquad \mu_0 = 313_{-10}^{+20} {\rm MeV}  
\label{eq:mu0_dis} 
\end{eqnarray} 
Although this seems a rather low scale, one may still hope that the
expansion parameter $\alpha (\mu_0) / 2 \pi \sim 0.34 \pm 0.04 $ makes
perturbation theory meaningful. A NLO analysis confirms this, at first
glance ``illegal'', expectation. This is the natural scale where {\it all
observables} are defined in the quark model. If we want to compute
observables at higher scales $\mu > \mu_0 $ one may use QCD
(perturbative) evolution using the quark model as an initial
condition. This way we generate the missing (perturbative) gluon
components explicitly in the form of QCD radiative corrections.

\section{THE SPECTRAL QUARK MODEL}

The subject of regularization in connection with high energy hadronic
properties in chiral quark models has always been tricky and
frustrating in the past, particularly when chiral and electromagnetic
invariance, anomalies and factorization were demanded simultaneously
(See Ref.~\cite{RuizArriola:2002wr} and references therein for a
comprehensive discussion). The SQM relies on a {\em spectral
regularization} of the chiral quark model proposed in
Ref.~\cite{RuizArriola:2001rr} and extensively developed in
Ref.~\cite{RuizArriola:2003bs} and is based on a generalized Lehmann
representation for the quark propagator\footnote{Note that an approach 
similar in spirit was described long ago by Efimov and Ivanov
~\cite{Efimov:1988yd}.}   
\begin{eqnarray}
S({p}) = \int_C d \omega { \rho( \omega ) \over \slashchar{p} - \omega
} = \int_C d \omega \frac{ \rho_V (\omega) \slashchar{p} +
\rho_S(\omega)\omega }{ p^2 - \omega^2 } = \frac{Z(p^2) } {\slashchar{p}
-M(p^2)},
\label{eq:spec0} 
\end{eqnarray}
where $C$ is a general contour to be determined. For obvious reasons
$\omega$ is called the spectral quark mass. Chiral and flavour gauge
invariance on the relevant vertex functions are obtained (up to
transverse terms) by means of the gauge technique of Delbourgo and
West~\cite{Delbourgo:1977jc} which is nothing but minimal coupling at
the spectral quark level. According to the analysis of
Refs.~\cite{RuizArriola:2001rr,RuizArriola:2003bs}, the proper
normalization and finiteness of hadronic observables are achieved by
requesting a set of {\em spectral conditions} for the moments of the
quark spectral function, $\rho(\omega)$, namely
\begin{eqnarray}
\rho_n && \equiv \int d\omega \omega^n \rho(\omega) = \delta_{n0}, 
 \;\;\; {\rm for} \; n=0,1,2,3,... \label{rhon}
\end{eqnarray} 
Physical observables turn out to be proportional to the inverse
moments,
\begin{eqnarray}
\rho_{-k} && \equiv \int d\omega \omega^{-k} \rho(\omega), \;\;\; {\rm
 for} \; k=1, 2, 3,... \label{rhoinv}
\end{eqnarray} 
as well as to the so called {\em ``log moments''},
\begin{eqnarray}
&& \rho'_n\equiv \int d\omega \log(\omega^2/\mu^2) \omega^n
\rho(\omega) \nonumber  =\int d\omega \log(\omega^2) \omega^n \rho(\omega), \;\;\; {\rm for} \; n=2,3,4,...  
\label{rholog}
\end{eqnarray} 
Note that the conditions (\ref{rhon}) remove the dependence on the
scale $\mu$ in (\ref{rholog}), thus we can drop it from the log. Using
these conditions one may prove a number of important features, like
fulfillment of chiral anomalies and factorization of form factors,
i.e. power like behaviour, in the high momentum limit. The main
assumption in Eq.~(\ref{eq:spec0}) is that of analyticity in the
complex plane but not of positivity, since then  even moments would not
vanish for a positive spectral function. Analyticity enables direct
calculations of hadron properties in the Minkowski space, an extremely
convenient circumstance when doing light-cone physics. The analyticity
assumption is also necessary in models formulated in Euclidean space,
although the practical continuation to Minkowski may become
numerically messy.

\begin{figure}
\begin{center}
\includegraphics[width=8cm]{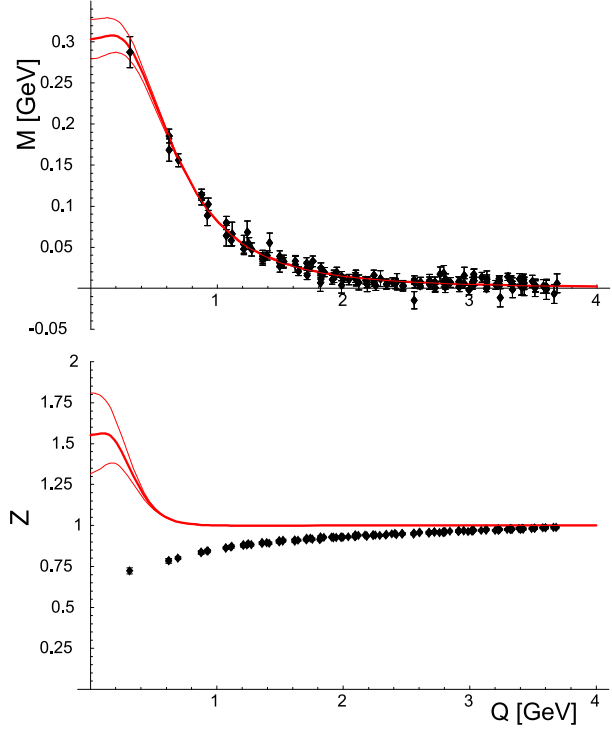}
\caption{The dependence of the quark mass $M$ (top) and the wave
function normalization $Z$ (bottom)  on the Euclidean momentum $Q$
obtained from the meson dominance model with $d_S=5/2$. Lattice data from Ref.~{\protect{\cite{Bowman:2002bm}}}.}
\label{fig:MZ}
\end{center}
\end{figure}

Although many results of our model do not depend on a particular
choice of functions fulfilling the spectral conditions (\ref{rhon}), a
particular realization embodies more predictions. The standard
approach to quark models is to assume a ``reasonable'' form for the
quark propagator and work out hadronic properties from there. This
poses the natural and difficult problem as to what is a reasonable
spectral function. Out of ignorance we take a radically different
approach, namely we deduce the spectral function from
phenomenology. Actually, $\rho_V(\omega) $, can be uniquely deduced
from the pion electromagnetic form factor, which for simplicity is
assumed to be of the vector dominance monopole form. The net
result~\cite{RuizArriola:2003bs} is the meson dominance (MD) model 
explicit realization which fulfills the above spectral conditions and
is given by
\begin{eqnarray}
\rho_V (\omega) &=& \frac{1}{2\pi i} \frac{1}{\omega}
\frac{1}{(1-4\omega^2/M_V^2)^{d_V}}, \qquad 
\rho_S (\omega)= \frac{1}{2\pi i} \frac{16(d_S-1)(d_S-2) \rho'_3}
{M_S^4 (1-4\omega^2/M_S^2)^{d_S}} , \label{rhos}
\end{eqnarray} 
The preferred values are $ d_V=d_S=5/2$ and $M_V=776 {\rm MeV}$
corresponds to the $\rho$-meson mass. The contour $C$ encircles the
branch cuts, {\em i.e.}, starts at $-\infty+i0$, goes around the
branch point at $-m_\rho/2$, and returns to $-\infty -i0$, with the
other section obtained by a reflexion with respect to the
origin~\cite{RuizArriola:2003bs}. Straightforward calculation yields
\begin{eqnarray} 
\frac{M(p^2)}{M(0)} &=& \frac{ 4 d_V p^2}{M_V^2}\frac{
\left(\frac{M_S^2}{M_S^2 -4 p^2} \right)^{d_S}}{
\left(\frac{M_V^2}{M_V^2 -4 p^2} \right)^{d_V} -1} 
\end{eqnarray} 
Remarkably, when $M(p^2)=p^2$ then $Z(p^2)=0 $ so that the quark
propagator has no pole. Instead it has cuts at $p^2 = M_V^2 /
4$. These feature is usually called analytic confinement and the
rationale for it has been described in detail in
Ref.~\cite{RuizArriola:2003bs}. A fit to the recent QCD lattice
simulation in Landau gauge~\cite{Bowman:2002bm} yields $ M (0) = 303
\pm 24~{\rm MeV}$ and $ M_S = 970 \pm 21~{\rm MeV}$, with the optimum
value of $\chi^2$ per degree of freedom equal to 0.72, yielding the
impressive agreement of $M(p^2)$ shown in the Fig.~(\ref{fig:MZ}). As
we see $Z(p^2)$ is not as good, so there is obviously still room for
improvement.

\newcommand{\MDM}{{\stackrel{\scriptstyle{\rm MD}} {=}}}

One of the main appeals of the SQM is its computational
simplicity. Using the above spectral conditions one can easily compute
many properties by doing standard one loop calculations. For instance,
the quark condensate for one flavour yields~\footnote{The upperscript
MD means evaluated in the Meson Dominance Model, Eq.~(\ref{rhos}).}. 
\begin{eqnarray}
 \langle \bar q q \rangle = - {N_c \over 4 \pi^2} \int d\omega
  \log(\omega^2) \omega ^3 \rho(\omega ) \MDM \frac{d_V M_S^4 N_c M(0)
  }{16 (d_S-1)(d_S-2) M_V^2 \pi^2 }.
\end{eqnarray}
With the MD spectral density, $\rho_S $, given by Eq.~(\ref{rhos})
this equation becomes an identity. The corresponding value of the
quark condensate is $\langle \bar q q \rangle =
-(243.0^{+0.1}_{-0.8}~{\rm MeV})^3 $ at the model scale $\mu_0 $,
which after LO evolution yields $\langle \bar q q \rangle =
-(301.0^{+0.1}_{-1.0}~{\rm MeV})^3 $ at $\mu=1 {\rm GeV}
$~\cite{RuizArriola:2002wr}. The vacuum energy density is scale
independent can be also be
computed to give
\begin{eqnarray}
B = - {N_c N_f \over 16 \pi^2} \int d\omega \log(\omega^2) \omega ^4
  \rho(\omega ) \MDM -\frac{N_c M_V^4}{64 \pi^2} &=&- (217~{\rm MeV}
  )^4 . \label{eq:w4log}
\end{eqnarray}
for three flavors, $N_f=3$. These values are reasonable according to
current QCD sum rules estimates $\langle \bar q q \rangle =
-(243.0~{\rm MeV})^3 $ (at $\mu = {\rm 1 GeV} ) $ and $ B = - (
224^{+35}_{-70} {\rm MeV}) ^4 $ (at any scale)~\cite{Ioffe:2002ee}.

\section{PION PROPERTIES} 

The pion weak decay constant can be worked out to give  
\begin{eqnarray} 
f_\pi^2 =- {N_c \over 4\pi^2 } \int d\omega \log (\omega^2 ) \omega^2
\rho(\omega)  \MDM  \frac{N_c M_V^2}{24\pi^2 } = ( 87 {\rm MeV})^2 
\label{eq:w2log}
\end{eqnarray} 
for $M_V = 776 {\rm MeV} $ in excellent agreement with value of
$f_\pi=88 {\rm MeV}$ in the chiral limit obtained from
ChPT~\cite{Gasser:1983yg}.  The pion charge form factor is given by
\begin{eqnarray}
F_V (t) \equiv F_\pi^{\rm em}(t)= 
-\frac{N_c}{4\pi^2 f_\pi^2} \int d\omega \rho(\omega) \omega^2 \int_0^1 dx
\log\left[ \omega ^2+x(1-x) t \right] \MDM \frac{M_V^2}{Q^2 + M_V^2} 
\end{eqnarray} 
with the built in vector meson dominance explicitly displayed. This
form describes the data remarkably well with $M_V \sim 730 {\rm MeV} $
up to $ Q^2 \sim 1.6 {\rm GeV}^2 $~\cite{Volmer:2000ek}.

\begin{figure}
\begin{center}
\includegraphics[width=8cm]{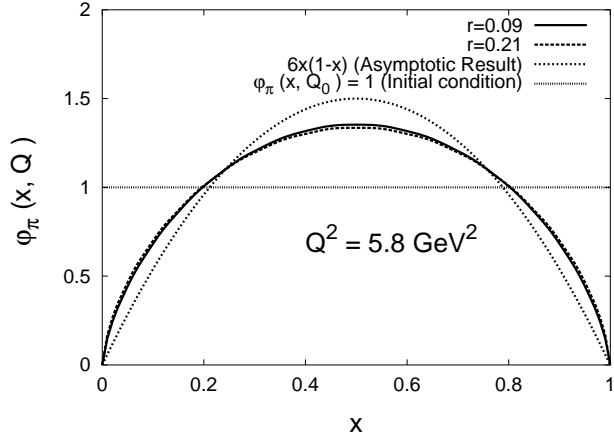}
\caption{{The pion distribution amplitude (PDA) in the chiral limit evolved
to the scale \protect{$Q^2 = (2.4 {\rm GeV})^2 $}.  The two values for
the evolution ratio $r=\alpha(Q) / \alpha(Q_0)$ reflect the
uncertainties based on an analysis~\cite{Bakulev:2002uc} 
of the CLEO data~\cite{CLEO98}.  We also show the unevolved PDA,
$\varphi_\pi(x,Q_0)=1$, and the asymptotic PDA, $\varphi_\pi
(x,\infty)=6x(1-x)$.}}
\label{fig:pda}
\end{center}
\end{figure}

A calculation along the lines of
Refs.~\cite{RuizArriola:2002bp,RuizArriola:2002wr} produces the
following identity between the light-cone pion wave function and the
unintegrated pion distribution function 
\begin{eqnarray}
\Psi(x,k_\perp)= q(x, k_\perp) = \frac{N_c}{4 \pi^3 f_\pi^2} \int
d\omega \rho(\omega) \frac{\omega^2}{k_\perp^2+\omega^2} \MDM
\frac{3M_V^3}{16\pi(k_\perp^2+M_V^2/4)^{5/2}}. \label{pionlcwf}
\end{eqnarray}
where we remind that the identity holds in our model at the low-energy
scale, $\mu_0$, of the model. In the MD the average transverse
momentum squared is equal to $\langle k_\perp^2 \rangle=M_V^2/2 =
(544~{\rm MeV})^2$. After integration over
$d^2 k_\perp$ the result for the PDA and PDF holds,   
\begin{eqnarray}
q(x) = \varphi (x) = 1 \qquad {\rm for} \qquad \mu = \mu_0  . 
\end{eqnarray}
The trend of both PDA and PDF being one has also been observed in
transverse lattice calculations~\cite{Dalley:2002nj}. The Gegenbauer
evolution taking into account the uncertainties discussed by
A. Bakulev in his talk~\cite{Bakulev:2002uc} produce the PDA depicted
in Fig.~(\ref{fig:pda}). In addition one may deduce after
DGLAP evolution of the PDF and Gegenbauer evolution of the
PDA the following relation
\begin{eqnarray}
\frac{\varphi_\pi (x, \mu) }{6 x (1-x) }-1 = \int_0^1 dy K(x,y) V_\pi
(y,\mu)  
\label{eq:int_eq} 
\end{eqnarray}  
where the explicit analytic expression for the scale independent
kernel, $K(x,y)$ is given in Ref.~\cite{RuizArriola:2002bp}. This is a
remarkable equation since it relates an inclusive process to an
exclusive one.  Using this equation one can regard the PDA measured at
CLEO as a prediction in terms of the PDF parameterizations of Durham
Ref.~\cite{Sutton:1991ay}. The QCD DGLAP-evolved PDF provide an
impressive description of the Durham
parameterization~\cite{Sutton:1991ay}. The comparison to Drell-Yan
data at $Q^2= 16 {\rm GeV}^2$ as well as the extension to non-skewed
generalized parton distributions can be found in the contribution by
WB in this workshop (see also Ref.~\cite{Broniowski:2003rp}). Using
Eq.~(\ref{eq:int_eq}) one gets the following estimate for the leading
twist contribution to the pion form factor at LO
\begin{eqnarray}
\frac{Q^2 F_{\gamma^* , \pi \gamma} (Q) }{2 f_\pi} \Big|_{\rm Twist-2} = 
\int_0^1 dx \frac{\varphi_\pi (x,Q )}{6x(1-x)} = 1.25 \pm 0.10  
\end{eqnarray}  
The experimental value obtained in CLEO~\cite{CLEO98} for the full
form factor is $ Q^2 F_{\gamma^* , \pi \gamma} (Q) / 2 f_\pi = 0.83
\pm 0.12 $ at $Q^2 = 5.8 {\rm GeV}^2 $. Considering that we have not
included neither NLO effects nor an estimate of higher twist
contributions, the 2-sigma discrepancy is not unexpected.  At
$k_\perp=0$ the following conditions are satisfied,
\begin{eqnarray}
\Psi(x,0) =\frac{N_c}{4 \pi^3 f_\pi^2} = q(x,0)\label{consist}.
\end{eqnarray}
The first equality corresponds to the proper normalization of the
anomalous decay $\pi^0 \to 2 \gamma $. Other anomalous vertices are
preserved as well~\cite{RuizArriola:2003bs}. 

To summarize, the SQM in conjunction with QCD evolution provides an
appealing and computationally cheap framework where the analyticity of
the quark propagator with a non-positive spectral function can be
exploited on the one hand and high energy QCD constraints can be
imposed.  Nonetheless, according to a revealing interview with
Rev. Matthews, a former Cannon of the Durham Cathedral, we should ask
not only {\it how} it works, but {\it why} !

\vskip.1cm
\noindent

\section*{ACKNOWLEDGEMENTS}

We would like to thank the organizers and particularly Simon Dalley for
the invitation.  This work is supported in part by funds provided by
the Spanish DGI with grant no. BFM2002-03218, and Junta de
Andaluc\'{\i}a grant no. FQM-225. Partial support from the Spanish
Ministerio de Asuntos Exteriores and the Polish State Committee for
Scientific Research, grant number 07/2001-2002 is also gratefully
acknowledged.

\end{document}